\documentclass[useAMS,usenatbib]{mn2e}

\usepackage{graphicx}	
\usepackage{amsmath}	
\usepackage{amssymb}	
\usepackage{upgreek}
\usepackage{morefloats}
\newcommand{\MS}{\ifmmode{\,}\else\thinspace\fi{\rm M}\ifmmode_{\odot}\else$_{\odot}$\fi}
\newcommand{\LS}{\ifmmode{\,}\else\thinspace\fi{\rm L}\ifmmode_{\odot}\else$_{\odot}$\fi}
\newcommand{\RS}{\ifmmode{\,}\else\thinspace\fi{\rm R}\ifmmode_{\odot}\else$_{\odot}$\fi}
\newcommand{\teff}{\ifmmode T_{\rm eff}\else$T_{\rm eff}$\fi}
\newcommand{\fo}{\ifmmode \nu_{\rm 1O}\else$\nu_{\rm 1O}$\fi}
\newcommand{\ff}{\ifmmode \nu_{\rm F}\else$\nu_{\rm F}$\fi}
\newcommand{\fb}{\ifmmode \nu_{\rm m}\else$\nu_{\rm m}$\fi}
\newcommand{\fu}{\ifmmode \nu_{\rm u}\else$\nu_{\rm u}$\fi}
\newcommand{\af}{\ifmmode A_{\rm F}\else$A_{\rm F}$\fi}
\newcommand{\aoaf}{\ifmmode A_{\rm 1O}/A_{\rm F}\else$A_{\rm 1O}/A_{\rm F}$\fi}
\newcommand{\Po}{\ifmmode P_{\rm 1O}\else$P_{\rm 1O}$\fi}
\newcommand{\Pf}{\ifmmode P_{\rm F}\else$P_{\rm f}$\fi}
\newcommand{\popf}{\ifmmode P_{\rm 1O}/P_{\rm F}\else$P_{\rm 1O}/P_{\rm F}$\fi}
\newcommand{\pol}{\frac{1}{2}}

\title[Anomalous Double-Mode RR Lyrae Stars]{Anomalous Double-Mode RR~Lyrae Stars in the Magellanic Clouds}
\author[Soszy{\'n}ski et al.]{I. Soszy{\'n}ski$^{1}$\thanks{E-mail: soszynsk@astrouw.edu.pl},
R. Smolec$^{2}$,
W. A. Dziembowski$^{1,2}$,
A. Udalski$^{1}$,
M. K. Szyma{\'n}ski$^{1}$,\and
{\L}. Wyrzykowski$^{1}$,
K. Ulaczyk$^{3}$,
R. Poleski$^{1,4}$,
P. Pietrukowicz$^{1}$,
S. Koz{\l}owski$^{1}$,\and
D. Skowron$^{1}$,
J. Skowron$^{1}$,
P. Mr{\'o}z$^{1}$,
\& M. Pawlak$^{1}$
\\
$^{1}$ Warsaw University Observatory, Al. Ujazdowskie 4, 00-478 Warszawa, Poland\\
$^{2}$ Nicolaus Copernicus Astronomical Center, Polish Academy of Sciences, ul. Bartycka 18, 00-716 Warszawa, Poland\\
$^{3}$ Department of Physics, University of Warwick, Gibbet Hill Road, Coventry, CV4 7AL, UK\\
$^{4}$ Department of Astronomy, Ohio State University, 140 W. 18th Ave., Columbus, OH 43210, USA\\
}

\begin{document}

\date{Accepted . Received ; in original form }

\pagerange{\pageref{firstpage}--\pageref{lastpage}} \pubyear{2015}

\maketitle

\label{firstpage}

\begin{abstract}
We report the discovery of a new subclass of double-mode RR~Lyrae stars in
the Large and Small Magellanic Clouds. The sample of 22 pulsating stars
have been extracted from the latest edition of the OGLE collection of
RR~Lyrae variables in the Magellanic System. The stars pulsating simultaneously
in the fundamental (F) and first-overtone (1O) modes have distinctly
different properties than regular double-mode RR~Lyrae variables (RRd
stars). The $P_\mathrm{1O}/P_\mathrm{F}$ period ratios of our anomalous RRd
stars are within a range 0.725--0.738, while ``classical'' double-mode
RR~Lyrae variables have period ratios in the range 0.742--0.748. In contrast to
the typical RRd stars, in the majority of the anomalous pulsators the
F-mode amplitudes are higher than the 1O-mode amplitudes. The light curves
associated with the F-mode in the anomalous RRd stars show different
morphology than the light curves of, both, regular RRd stars and
single-mode RRab stars. Most of the anomalous double-mode stars show
long-term modulations of the amplitudes (Blazhko-like effect).
Translating the period ratios into the abundance parameter, $Z$, we find
for our stars $Z\in(0.002,0.005)$ -- an order of magnitude higher values
than typical for RR~Lyrae stars. The mass range of the RRd stars inferred
from the $W_I$ vs. $\Pf$ diagram  is $(0.55-0.75)\MS$. These parameters
cannot be accounted for with single star evolution assuming a Reimers-like
mass loss. Much greater mass loss caused by interaction with other stars is
postulated. We blame the peculiar pulsation properties of our stars to the
parametric resonance instability of the 1O-mode to excitation of the F- and
2O-modes as with the inferred parameters of the stars $2\omega_{\rm
1O}\approx\omega_{\rm F}+\omega_{\rm 2O}$.
\end{abstract}


\section{Introduction}

The first RR~Lyrae star pulsating simultaneously in two radial modes (an
RRd star) -- AQ Leonis -- was identified by \cite{jer77}. For years, RRd
stars have been considered as pure radial pulsators, with no peculiar
behaviors like non-radial oscillations or Blazhko effect. This simple
picture has changed in recent years with the advent of space-based and
massive ground-based photometry. Observations from the space telescopes --
MOST, CoRoT, and Kepler \citep[e.g.][respectively]{gru07,cha12,mol15} --
led to the discovery of additional non-radial modes in RRd variables.
Similar detections were also done in the ground-based photometry
\citep{net15,jur15}. See \cite{kur16} for the most recent summary (their
table 4) and \cite{dzi16} for the theoretical model explaining the
excitation of non-radial modes in RR Lyrae stars.

\begin{figure*}
\includegraphics[width=15cm]{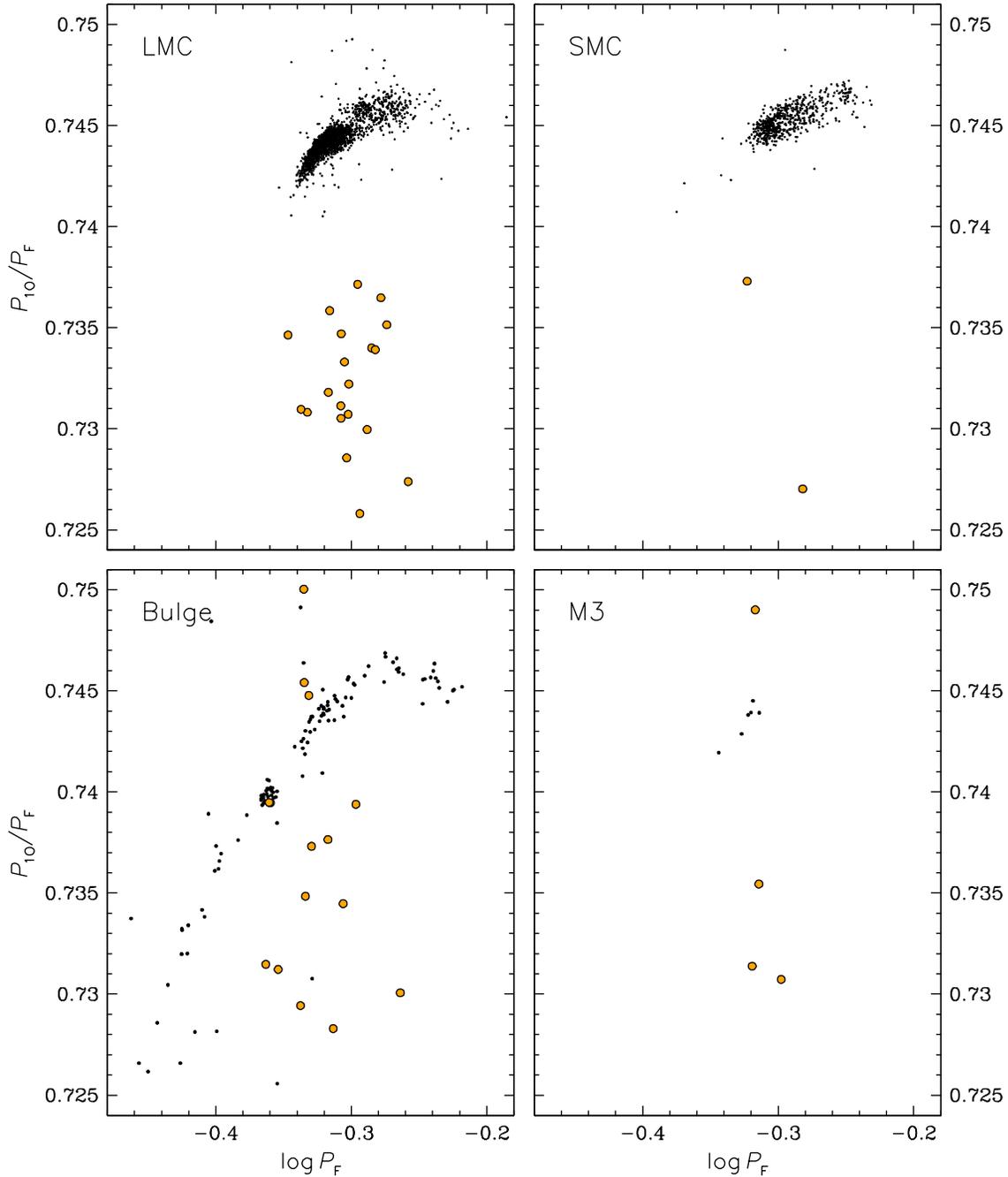}
\caption{Petersen diagrams for RRd stars in four different stellar systems:
LMC, SMC \citep{sos14}, Galactic bulge \citep{sos14,smo15}, and globular
cluster M3 \citep{jur14,jur15}. Small black dots mark the ``classical'' RRd
stars, large yellow circles mark the anomalous RRd stars in the Magellanic
Clouds or Blazhko RRd stars in the Galactic bulge and M3.}
\label{fig:pet}
\end{figure*}

Nowadays, the number of known RRd stars reaches several thousands and most of them have
been discovered in the time-series photometric databases obtained by the
Optical Gravitational Lensing Experiment (OGLE). The Large (LMC) and Small
Magellanic Clouds (SMC) host particularly rich populations of double-mode
RR~Lyrae stars. The latest release of the OGLE Collection of Variable Stars
\citep{sos16} contains in total 2624 double-mode RR~Lyrae stars in
the Magellanic Clouds. RRd stars constitute about 5\% and 10\% of all
RR~Lyrae variables in the LMC and SMC, respectively. In turn, in the central
regions of the Milky Way the rate of double-mode RR~Lyrae is below 0.5\% of
the entire sample \citep{sos11,sos14}.

\begin{figure*}
\centering
\includegraphics[width=17.0cm]{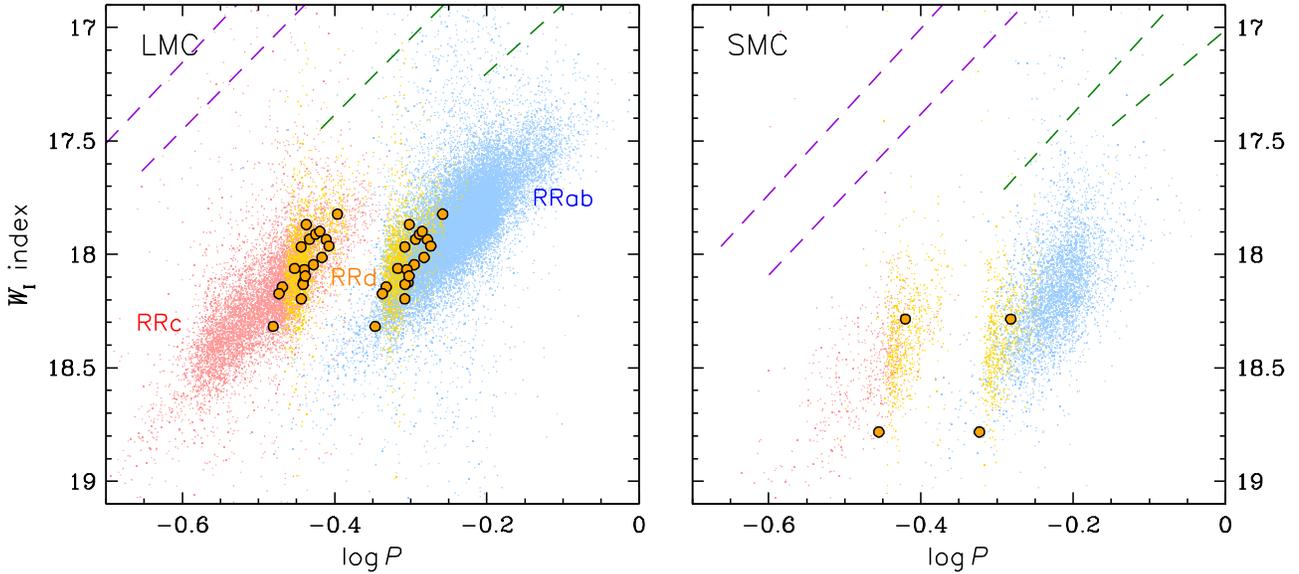}
\caption{Period--luminosity diagrams for RR~Lyrae stars in the LMC (left
panel) and SMC (right panel). Blue, red, and yellow dots mark RRab, RRc,
and classical RRd stars, respectively. Large circles mark the anomalous
RRd stars. Purple and green dashed lines indicate the fits the
period--luminosity relations for classical Cepheids (first-overtone and
second-overtone modes) and anomalous Cepheids (fundamental and
first-overtone modes).}
\label{fig:pl}
\end{figure*}

Double-mode RR~Lyrae variables occupy a limited region in the Petersen
diagram in which the ratios of the first-overtone to fundamental-mode
periods ($P_\mathrm{1O}/P_\mathrm{F}$) are plotted against the logarithm of
the fundamental-mode period ($P_\mathrm{F}$). In Fig.~\ref{fig:pet} we
present Petersen diagrams for RRd stars in four environments: LMC, SMC
\citep{sos16}, Galactic bulge \citep{sos14}, and globular
cluster M3 \citep{jur14,jur15}. Most of the RRd stars form a curved
sequence continuing over a narrow range of period ratios
$0.742<P_\mathrm{1O}/P_\mathrm{F}<0.748$. The only known exception to this
rule is the Galactic bulge, where RRd variables form a sequence reaching
the period ratios as small as 0.726, which can be explained by a high metal
abundance in these stars \citep{sos11}.

In many environments, a number of double-mode RR~Lyrae stars lies below or
above the curved sequence in the Petersen diagram. The first RRd stars
showing the Blazhko effect (quasi-periodic modulations of pulsation
amplitudes and/or phases) were found just among these outliers in the
Galactic bulge \citep{sos14,smo15}. The same feature was
discovered by \cite{jur14} for four RRd stars in the globular cluster
M3. It seems that Blazhko RRd stars avoid the sequence delineated by
regular RRd variables in the Petersen diagram (bottom panels of
Fig.~\ref{fig:pet}).

In this paper, we report the discovery of a group of 22 double-mode RR~Lyrae
stars in the Magellanic Cloud. Our variables have period ratios within the
range $0.725<P_\mathrm{1O}/P_\mathrm{F}<0.738$ -- considerably lower than
observed in the ``classical'' RRd stars in these galaxies. We show that
both classes of double-mode pulsators present distinct features, like
amplitude ratios, shapes of the light curves, presence of the Blazhko
effect, and the spatial distribution on the sky. We call the newly
identified variables ``anomalous RRd stars''.

\section{Selection of Anomalous RRd Stars}

The sample of anomalous RRd stars has been extracted from the OGLE
collection of $45\;451$ RR~Lyrae variables in the Magellanic Clouds
\citep{sos16}. Each light curve was searched for secondary periodicities
and, as a result, more than 2600 ``classical'' RRd stars have been
identified. Moreover, five RR~Lyrae stars with superimposed eclipsing
variability and a number of other multi-periodic variables were found. In
the latter group we noticed 22 (20 in the LMC and 2 in the SMC) double-mode
RR~Lyrae stars located just below ``classical'' RRd stars in the Petersen
diagram (Fig.~\ref{fig:pet}, top panels). With the longer periods ranged
between 0.45~d and 0.55~d and the period ratios between 0.725 and 0.738,
these stars distinctly cluster in this region of the Petersen diagram, but
they do not form a continuous structure with typical RRd stars.

\begin{figure*}
\includegraphics[width=17.0cm]{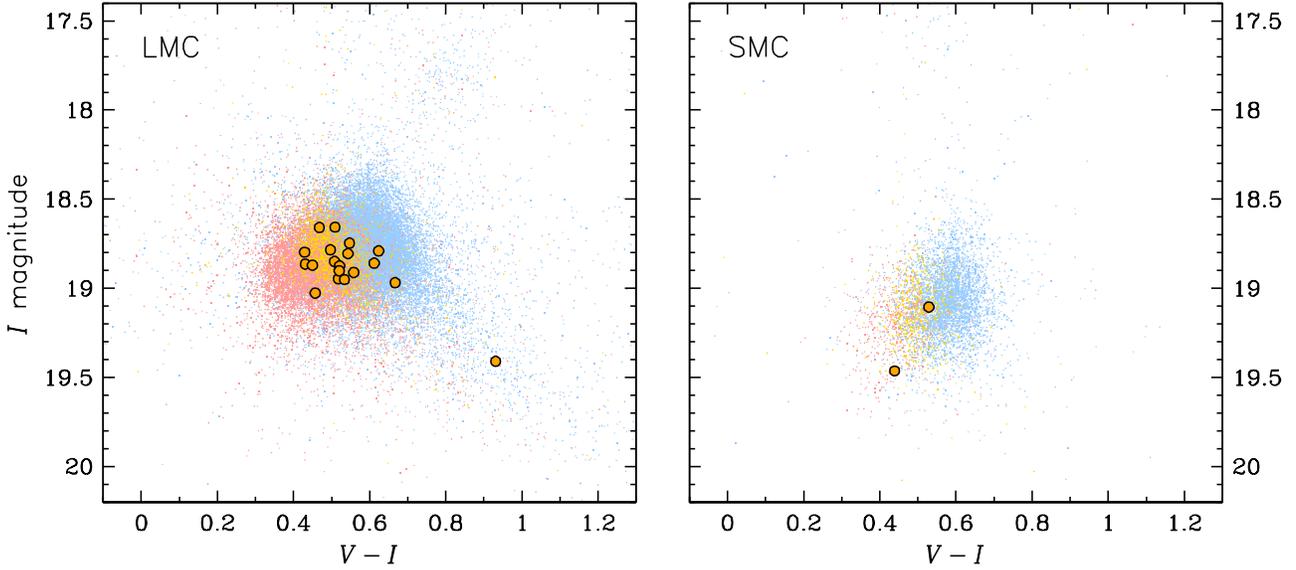}
\caption{Colour--magnitude diagrams for RR~Lyrae stars in the Magellanic
Clouds. The colour symbols are the same as in Fig.~\ref{fig:pl}.}
\label{fig:cmd}
\end{figure*}

In the first place, we ensured that our nonstandard double-periodic
variables are RR~Lyrae stars. Among radially pulsating stars in the
Magellanic Clouds, double- or triple-mode classical Cepheids may have
periods around 0.5~d \citep[e.g.][]{sos15b}, but generally they are
brighter than RR~Lyrae stars with the same periods. Anomalous Cepheids are
fainter than classical Cepheids \citep{sos15a}, but these stars are
exclusively single-mode pulsators. In Fig.~\ref{fig:pl}, we plot
period--luminosity diagram (where as a luminosity we use the
extinction-independent Wesenheit index, defined as $W_I=I-1.55(V-I)$), for
RR~Lyrae stars in the LMC and SMC. Both modes of the anomalous double-mode
variables are overplotted with large yellow circles. Also the fits to the
period--luminosity relations for classical and anomalous Cepheids are
displayed in Fig.~\ref{fig:pl}. One can see that our double-mode stars lie
exactly on the period--luminosity relations followed by fundamental-mode
(RRab) and first-overtone (RRc) RR~Lyrae stars. Classical and anomalous
Cepheids are much brighter than RR~Lyrae stars with the same periods. The
$(V-I)$ colours of our variables are practically the same as for typical
RRd stars (Fig.~\ref{fig:cmd}). Only one star in the LMC is highly reddened
by the interstellar extinction and one object in the SMC seems to be
slightly bluer and fainter than typical double-mode RR~Lyrae stars, but still
in the range occupied by RR~Lyrae stars. We conclude that our targets must be
RR~Lyrae stars or, at least, are photometrically indistinguishable from
RR~Lyrae stars. The positions of these objects in the period--luminosity
diagram show that they have fundamental and first-overtone radial modes
simultaneously excited.

\begin{figure*}
\centering
\includegraphics[width=17.6cm]{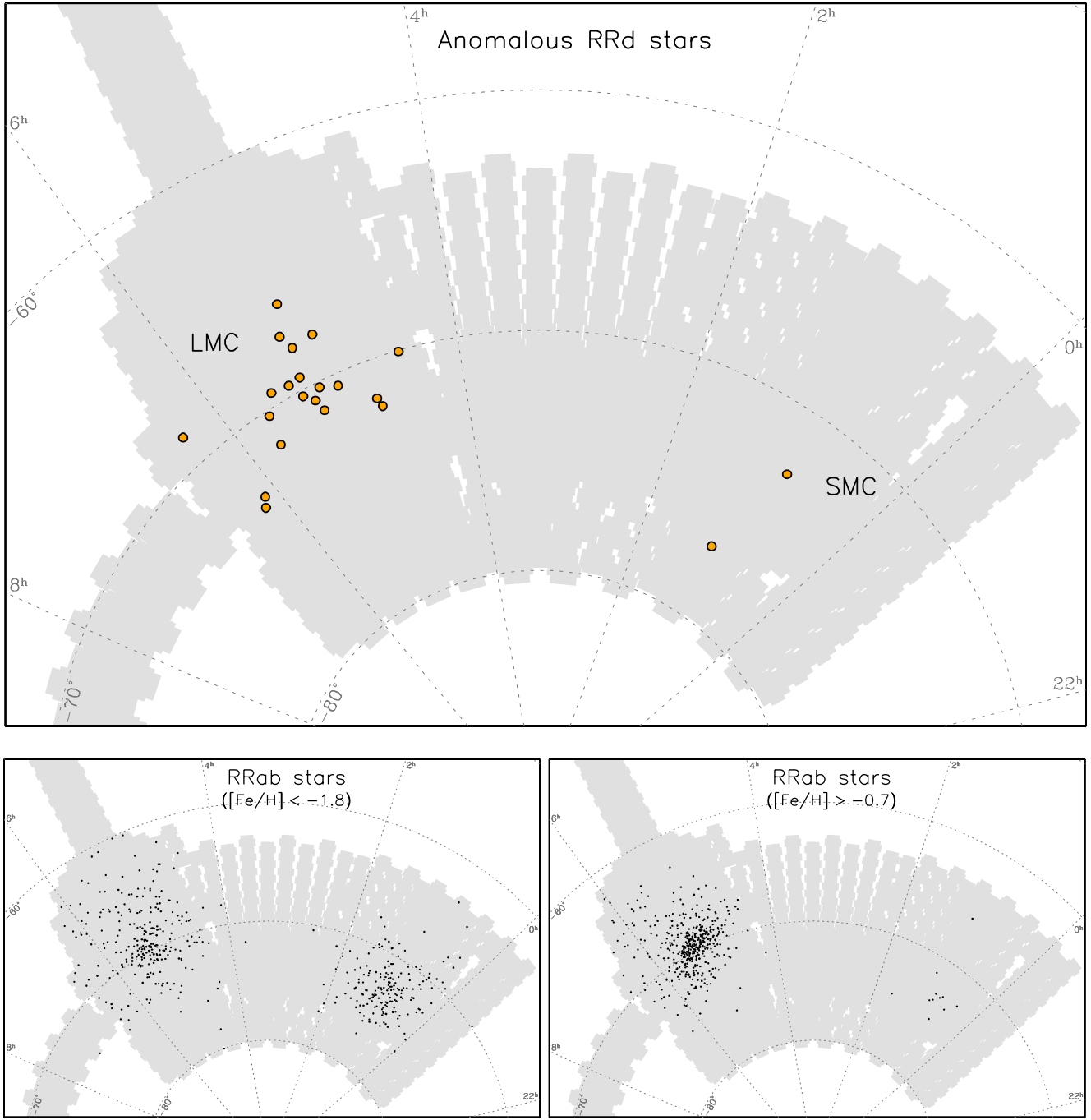}
\caption{Spatial distribution of RR~Lyrae stars in the Magellanic
System. Upper panel shows the positions of anomalous RRd stars in the
sky. In the bottom panels distributions of metal poor
($\mathrm{[Fe/H]}<-1.8$) and metal rich ($\mathrm{[Fe/H]}>-0.7$) RRab
stars are presented. Gray area shows the sky coverage of the OGLE
fields.}
\label{fig:map}
\end{figure*}

Positions of our anomalous RRd stars in the sky are presented in
Fig.~\ref{fig:map}. For comparison, in the bottom panels we plot the
distributions of metal poor ($\mathrm{[Fe/H]}<-1.8$) and metal rich
($\mathrm{[Fe/H]}>-0.7$) RRab stars. The metallicities were derived using
the photometric method developed by \cite{jur96} and calibrated to the $I$
band by \cite{smo05}. The anomalous RRd stars tend to cluster towards the
center of the LMC, confirming their genuine LMC membership and indicating
likely higher metallicity and younger population (see later in the
Discussion). We note that metallicity determination for anomalous RRd stars
using the \cite{jur96} method is not possible, due to the anomalous shape of
the light curves, as we describe in more detail below.

\section{Properties of anomalous RRd stars}

The basic properties of the anomalous RRd stars are collected in
Tab.~\ref{tab:basics}. These are: periods of the two pulsation modes,
period ratio, amplitude of the fundamental mode, amplitude ratio, and mean
$I$ and $V$ band magnitudes. The complete solution of the light curve
decomposition is provided in the Appendix available in the on-line
version of the journal. Tables A1-A22 contain identifications of all
detected frequencies, their values, amplitudes and phases, all with
standard errors. Because the vast majority of the OGLE observations have
been obtained in the {\it I} photometric band, our analysis is based on
the light curves in this passband. Between March 2010 and July 2015 OGLE
collected from about 200 to nearly 750 {\it I}-band data points per star.
A median uncertainty of individual photometric measurements for stars
as faint as RR Lyrae variables in the Magellanic Clouds is about 0.04~mag.

Anomalous RRd stars differ from normal RRd
variables not only by the $\popf$ period ratios. The first striking
difference is the amplitude ratios of both modes. In the bulk of typical
RRd stars, the first overtone is the dominant pulsation mode. In anomalous
RRd stars in the Magellanic Clouds, the fundamental mode has larger
amplitude than the overtone mode in all but three variables.

\begin{figure*}
\centering
\includegraphics[width=12.0cm]{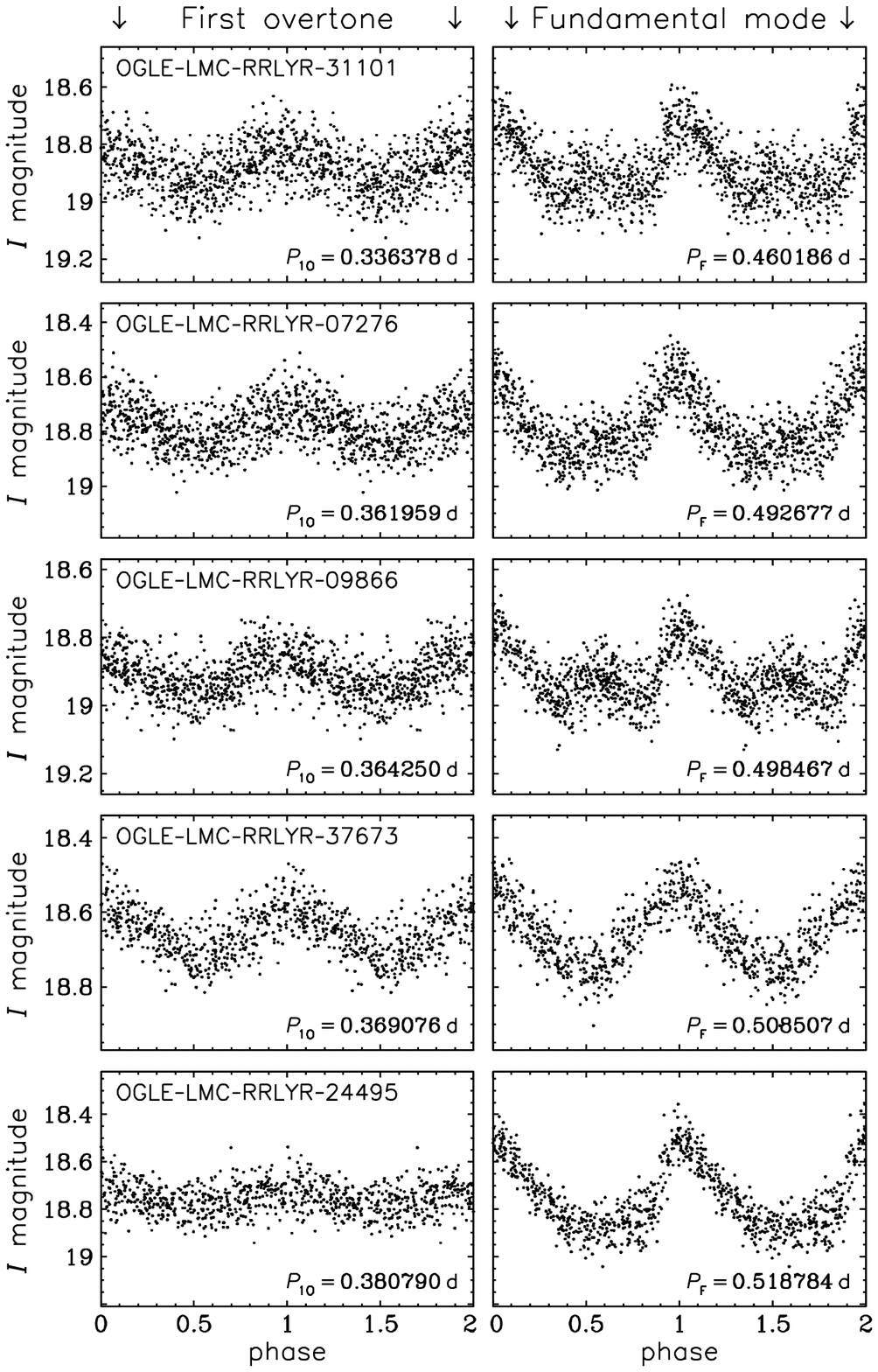}
\caption{Light curves for a sample of anomalous RRd stars decomposed into
the first overtone (left panels) and fundamental mode (right panels)
variations.}
\label{fig:lcurves}
\end{figure*}

\begin{table*}
\caption{Basic properties of anomalous RRd stars: star's ID, 1O and F mode
periods, $\popf$ period ratio, Fourier amplitude of the fundamental mode
and $\aoaf$ amplitude ratio, mean $I$- and $V$-band magnitudes.}
\label{tab:basics}
\begin{tabular}{lrrrrrrr}
Star's ID & $\Po$\thinspace (d) & $\Pf$\thinspace (d) & $\popf$ & $A_{\rm F}$\thinspace (mag) & $\aoaf$ & $\langle I\rangle$(mag) &  $\langle V\rangle$(mag) \\
\hline
OGLE-LMC-RRLYR-00013  &  0.363332  &  0.495475  &  0.7333  &  0.088  &  0.68  &  18.874  &  19.395 \\
OGLE-LMC-RRLYR-03878  &  0.391356  &  0.532351  &  0.7351  &  0.061  &  1.11  &  18.805  &  19.348 \\
OGLE-LMC-RRLYR-04176  &  0.373427  &  0.506581  &  0.7372  &  0.149  &  0.25  &  18.910  &  19.468 \\
OGLE-LMC-RRLYR-07276  &  0.361959  &  0.492677  &  0.7347  &  0.111  &  0.54  &  18.797  &  19.226 \\
OGLE-LMC-RRLYR-08767  &  0.383104  &  0.522000  &  0.7339  &  0.104  &  0.58  &  18.784  &  19.281 \\
OGLE-LMC-RRLYR-08917  &  0.362271  &  0.497235  &  0.7286  &  0.089  &  0.65  &  18.951  &  19.485 \\
OGLE-LMC-RRLYR-09866  &  0.364250  &  0.498467  &  0.7307  &  0.063  &  0.87  &  18.901  &  19.421 \\
OGLE-LMC-RRLYR-10802  &  0.365461  &  0.499116  &  0.7322  &  0.089  &  0.18  &  18.657  &  19.166 \\
OGLE-LMC-RRLYR-12487  &  0.330622  &  0.450038  &  0.7347  &  0.165  &  0.27  &  19.026  &  19.483 \\
OGLE-LMC-RRLYR-14584  &  0.355491  &  0.483089  &  0.7359  &  0.160  &  0.36  &  19.147  &   --~~~ \\
OGLE-LMC-RRLYR-19077  &  0.359980  &  0.492359  &  0.7311  &  0.076  &  0.78  &  18.864  &  19.295 \\
OGLE-LMC-RRLYR-21363  &  0.359766  &  0.492473  &  0.7305  &  0.044  &  2.14  &  19.409  &  20.340 \\
OGLE-LMC-RRLYR-22167  &  0.375823  &  0.514847  &  0.7300  &  0.051  &  1.49  &  18.859  &  19.471 \\
OGLE-LMC-RRLYR-24495  &  0.380790  &  0.518784  &  0.7340  &  0.169  &  0.19  &  18.746  &  19.293 \\
OGLE-LMC-RRLYR-27200  &  0.401566  &  0.552068  &  0.7274  &  0.105  &  0.23  &  18.790  &  19.414 \\
OGLE-LMC-RRLYR-27329  &  0.388186  &  0.527078  &  0.7365  &  0.097  &  0.63  &  18.968  &  19.635 \\
OGLE-LMC-RRLYR-30248  &  0.339935  &  0.465144  &  0.7308  &  0.143  &  0.22  &  18.947  &  19.465 \\
OGLE-LMC-RRLYR-31101  &  0.336378  &  0.460186  &  0.7310  &  0.096  &  0.71  &  18.870  &  19.320 \\
OGLE-LMC-RRLYR-37673  &  0.369076  &  0.508507  &  0.7258  &  0.103  &  0.64  &  18.659  &  19.127 \\
OGLE-LMC-RRLYR-38132  &  0.352750  &  0.482028  &  0.7318  &  0.093  &  0.89  &  18.849  &  19.357 \\
OGLE-SMC-RRLYR-1125   &  0.350540  &  0.475437  &  0.7373  &  0.166  &  0.29  &  19.463  &  19.902 \\
OGLE-SMC-RRLYR-4726   &  0.379910  &  0.522562  &  0.7270  &  0.188  &  0.18  &  19.105  &  19.634 \\
\hline
\end{tabular}
\end{table*}

\begin{figure*}
\includegraphics[width=17.0cm]{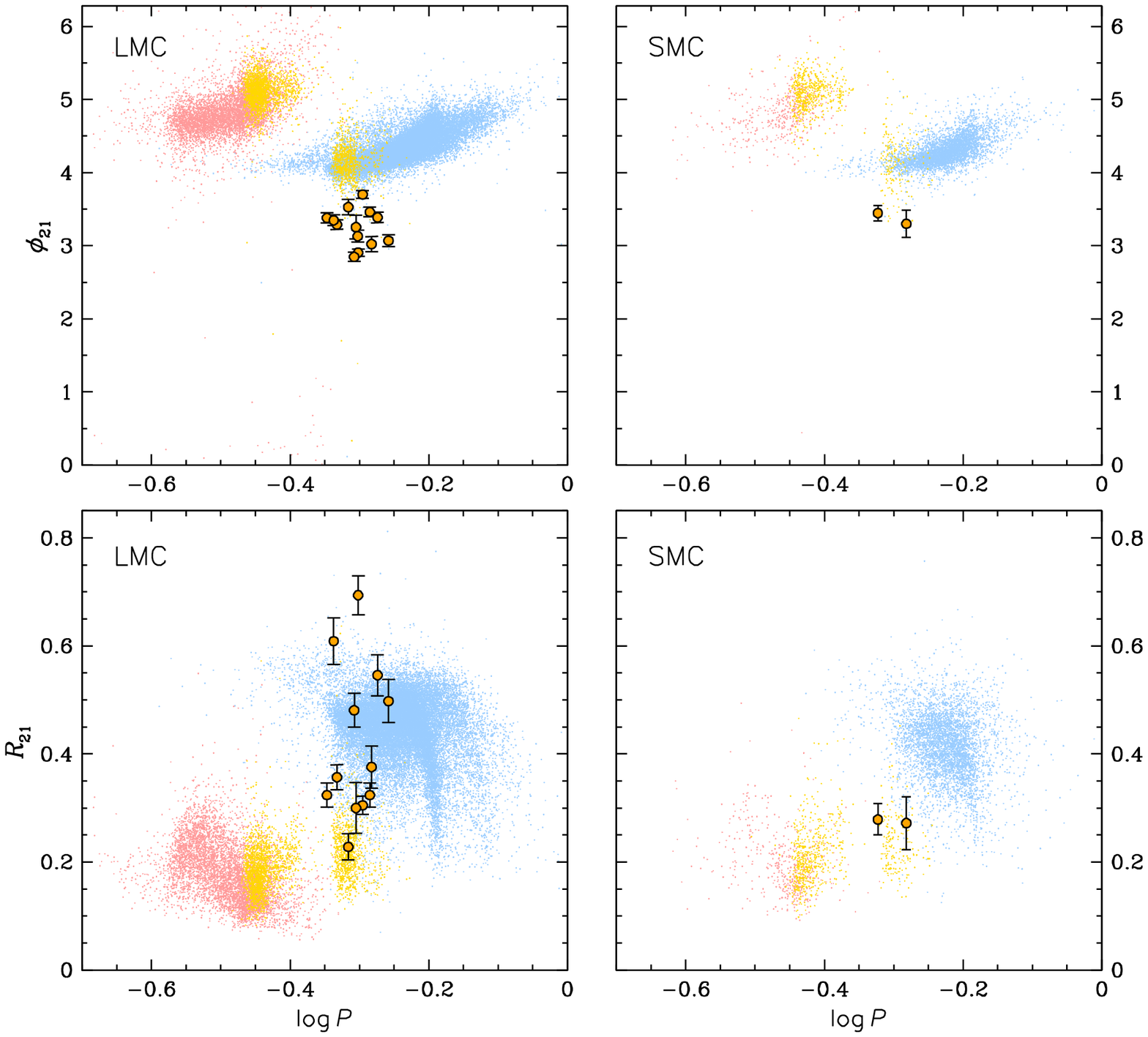}
\caption{Fourier coefficients, $\phi_{21}$ vs. $\log P$ (top panels)
and $R_{21}$ vs. $\log P$ (bottom panels) for RR~Lyrae stars in the Magellanic
system. Symbols are the same as in Fig.~\ref{fig:pl}.}
\label{fig:fparams}
\end{figure*}

Moreover, anomalous RRd stars exhibit distinct features of the light curves
morphology. Fig.~\ref{fig:lcurves} displays light curves of five
variables. For each star we present the two modes, prewhitened for the
other mode. The first-overtone components have nearly sinusoidal shapes,
while the fundamental-mode light curves are usually asymmetric, but they
are significantly different than their single-mode counterparts (RRab
stars). It can be quantitatively shown using the coefficients of the
Fourier light curve decomposition of the form:
$$I(t)=A_0+\sum_{i=1}^{N}  A_{i}\cos(i \omega t+\phi_{i})\,,$$
where $I(t)$ is the observed {\it I}-band magnitude at time $t$,
$\omega=2\pi/P$ is the angular frequency, $P$ is the pulsation period in
days, $A_i$ and $\phi_{i}$ represent the amplitude and phase-shift for
$i$th-order respectively, and $N$ is the optimum order of the fit adopted
individually for each light curve. The most telling is the
period--$\phi_{21}$ diagram ($\phi_{21}=\phi_2-2\phi_1$), which we present
in the top panels of Fig.~\ref{fig:fparams}. The Fourier phases for the fundamental mode
variation in anomalous RRd stars are systematically lower than in both RRab
and classical RRd stars. The amplitude ratios, $R_{21}=A_2/A_1$ (bottom panels of Fig.~\ref{fig:fparams}), on the
other hand, cover a wide range and can exceed the typical values observed
in RRab and classical RRd stars, a manifestation of the peculiar light
curve shape, with pronounced bump often present at the minimum brightness
(Fig.~\ref{fig:lcurves}). Similar analysis is not possible for the first
overtone, as only in a few cases the harmonic is significant.

Another difference is the presence of a modulation of pulsation (the
Blazhko effect). Modulation of pulsation amplitude and/or phase manifests
through the presence of equidistant multiplet structures in the frequency
spectrum, centred on the radial mode frequency. Frequency separation within
the multiplet corresponds to the modulation frequency, $\fb$, and its
inverse corresponds to the modulation period. The modulation properties of
anomalous RRd stars, determined from the analysis of their frequency
spectra, are summarised in Tab.~\ref{tab:mod}, in which we list which
radial mode is modulated, provide the modulation period, and modulation
amplitudes of both modes. Modulation amplitude is defined as the ratio of
the amplitude of the highest side peak at radial mode frequency to the
amplitude of the radial mode. The last column of Tab.~\ref{tab:mod} lists
all detected modulation side peaks. Signature of modulation is detected in
18 out of 22 anomalous RRd stars. Only in six stars we detect equidistant
triplet, however. In the majority of cases, only one side peak is detected,
which is also interpreted as due to modulation. The modulation multiplets
may be highly asymmetric, provided amplitude and phase modulation are of
comparable strength and are shifted in phase by $\approx\pm\pi/2$
\citep{ben11}. Interestingly, at the frequency of the fundamental mode, it
is the lower frequency side peak that is nearly always the higher one (for
stars with triplet structure) or it is the only modulation side peak that
is detected. The only exception is OGLE-LMC-RRLYR-08917 which we discuss in
more detail below. We also note that the amplitude of the modulation of the
fundamental mode is typically very high. In two cases (OGLE-LMC-RRLYR-00013
and OGLE-LMC-RRLYR-22167) the low-frequency modulation side peak,
$\ff-\fb$ is slightly higher than the peak at the fundamental mode
frequency, $\ff$ (Tab.~\ref{tab:mod}). In fact, other interpretation of the
close peaks detected in the vicinity of the fundamental mode is
possible. The highest peak, of lower frequency, may correspond to the
fundamental mode ($\ff-\fb=\ff'$) and the other peak may correspond to the
modulation side peak ($\ff=\ff'+\fb$). Such solution seems less
reasonable, however, at least in the case of OGLE-LMC-RRLYR-22167, as
instead of symmetric triplet centered at $\ff$ (see last column of
Tab.~\ref{tab:mod}), asymmetric quintuplet appears at $\ff'$.

\begin{table*}
\caption{Modulation properties of the anomalous RRd stars: star's ID, which
mode is modulated, modulation period, modulation amplitude and list of
all modulation side peaks detected in the frequency spectrum (bold face
font indicates the highest side peak at radial mode frequency, asterisk
indicates remnant power after the prewhitening and question mark
indicates weak detection with $3.0<{\rm S/N}<4.0$).}
\label{tab:mod}
\begin{tabular}{lrrrrl}
\hline
Star's ID &  mode & $P_{\rm m}$ & $A_{\rm m}$(F) & $A_{\rm m}$(1O) & detected modulation components\\
 &  & (d) & (mag) & (mag) & \\
\hline
OGLE-LMC-RRLYR-00013 & F/1O & 123.8(3)   & 1.05 & 0.49 & $\boldsymbol{\ff-\fb}$, $\boldsymbol{\fo-\fb}$ \\
OGLE-LMC-RRLYR-03878 &   F  & 201.5(9)   & 0.95 &      & $\boldsymbol{\ff-\fb}$  \\
OGLE-LMC-RRLYR-04176 &   F  &  19.72(1)  & 0.26 &      & $\boldsymbol{\ff-\fb}$ \\
OGLE-LMC-RRLYR-07276 &   F  &  44.91(4)  & 0.52 &      & $\boldsymbol{\ff-\fb}$*, $\ff+\fb$, $\ff+\fo-\fb$, $\ff+\fo+\fb$ \\
OGLE-LMC-RRLYR-08767 &   F  & 223.4(1.3) & 0.46 &      & $\boldsymbol{\ff-\fb}$    \\
OGLE-LMC-RRLYR-08917 & F/1O & 163.8(2)   & 0.64 & 1.24 & $\boldsymbol{\ff+2\fb}$, $2\ff+2\fb$, $\boldsymbol{\fo-\fb}$, $\fo+\fb$, $\ff+\fo-\fb$, \\
\multicolumn{5}{c}{} & $\ff+\fo+\fb$\\
OGLE-LMC-RRLYR-09866 &   F  & 202.3(1.4) & 0.81 &      & $\boldsymbol{\ff-\fb}$, $\ff+\fb$, $\fo-\ff-\fb$(?) \\
OGLE-LMC-RRLYR-10802 &   F  &  51.12(4)  & 0.67 &      & $\boldsymbol{\ff-\fb}$*, $\fb$ \\
OGLE-LMC-RRLYR-12487 & F/1O & 130.6(6)   & 0.19 & 0.44 & $\boldsymbol{\ff-\fb}$, $\boldsymbol{\fo+\fb}$ \\
OGLE-LMC-RRLYR-14584 & F/1O & 204.9(9)   & 0.35 & 0.91 & $\boldsymbol{\ff-\fb}$, $\ff+\fb$, $\boldsymbol{\fo+\fb}$  \\
OGLE-LMC-RRLYR-19077 &   1O &  56.4(1)   &      & 0.74 & $\boldsymbol{\fo-\fb}$, $\ff+\fo-\fb$ \\
OGLE-LMC-RRLYR-22167 & F/1O & 143.3(3)   & 1.09 & 0.21 & $\boldsymbol{\ff-\fb}$, $\ff+\fb$, $\ff+\fo-\fb$, $\boldsymbol{\fo-\fb}$, $\fo-2\fb$(?) \\
OGLE-LMC-RRLYR-24495 &   F  & 174.3(8)   & 0.23 &      & $\boldsymbol{\ff-\fb}$, $\ff+\fo-\fb$, $2\ff+\fo-\fb$, $3\ff+\fo-\fb$ \\
OGLE-LMC-RRLYR-27200 &   F  & 201.7(7)   & 0.52 &      & $\boldsymbol{\ff-\fb}$, $\ff+\fb$(?)\\
OGLE-LMC-RRLYR-27329 &   F  &  44.86(9)  & 0.42 &      & $\boldsymbol{\ff-\fb}$* \\
OGLE-LMC-RRLYR-31101 & F/1O & 206.2(7)   & 0.60 & 0.45 & $\boldsymbol{\ff-\fb}$*, $\boldsymbol{\fo-2\fb}$ \\
OGLE-SMC-RRLYR-1125  &   F  &  74.5(4)   & 0.19 &      & $\boldsymbol{\ff-\fb}$  \\
OGLE-SMC-RRLYR-4726  &   F  & 144.5(1.3) & 0.35 &      & $\boldsymbol{\ff-\fb}$  \\
\hline
\end{tabular}
\end{table*}

Modulation periods in the anomalous RRd stars are in the $\sim
20-220$\thinspace d range. In 11 stars only the fundamental mode is
modulated. In one star the first overtone is the only modulated mode and in
the remaining six stars both radial modes are modulated with the common
period. The only exceptions are OGLE-LMC-RRLYR-08917 and
OGLE-LMC-RRLYR-31101.

In OGLE-LMC-RRLYR-31101, at the frequency of the first overtone, we detect
only one side peak at $\fo-2\fb$. Two interpretations are possible. Either
we detect an incomplete quintuplet structure, or first overtone is
modulated with period twice as short as the fundamental mode.

OGLE-LMC-RRLYR-08917 is even more intriguing; the most reliable
interpretation of the detected modulation side peaks is presented in the
last column of Tab.~\ref{tab:mod}. The equidistant modulation triplets are
centred at $\fo$ and at the combination frequency, $\fo+\ff$. Then, the
only side peak at $\ff$ is located at $\ff+2\fb$ indicating that
fundamental mode is modulated with period twice as short as the first
overtone.

In some cases, after the prewhitening, remnant unresolved power is present
at the modulation side peaks (marked with asterisk in the last column of
Tab.~\ref{tab:mod}). It indicates that the modulations are irregular.

\section{Nature and cause of the anomalous RRd pulsation}

To get more insight into the nature of anomalous RRd stars we have computed
a grid of linear pulsation models. We used a hybrid of our Warsaw codes:
the updated version of the \cite{dzi77} code and the codes by
\cite{smo08}. With the former code, very precise computation of mode
periods, including non-radial modes, is possible, but, since convection is
treated in the frozen-in approximation, the models do not predict the red
edge of the instability strip (IS). This is possible with the latter code,
which implements the convection model by \cite{kuh86}. The two codes use
exactly the same OPAL \citep{igl96} opacity tables and adopt a solar mixture
as in \cite{asp09}. The two codes were calibrated to produce the same
location of the blue edge of the IS in the HR diagram by suitable
adjustment of the mixing-length parameter in the Dziembowski's (1977)
code. We use unfitted envelope models. We focus on the more numerous LMC
sample of anomalous RRd stars. For a given model, mean colours and
Wesenheit index were computed through interpolation in the
\cite{kur05}\footnote{http://kurucz.harvard.edu/} atmosphere models and
adopting a 18.5~mag distance modulus to the LMC \citep{pie13}.

\begin{figure*}
\centering
\resizebox{\hsize}{!}{\includegraphics{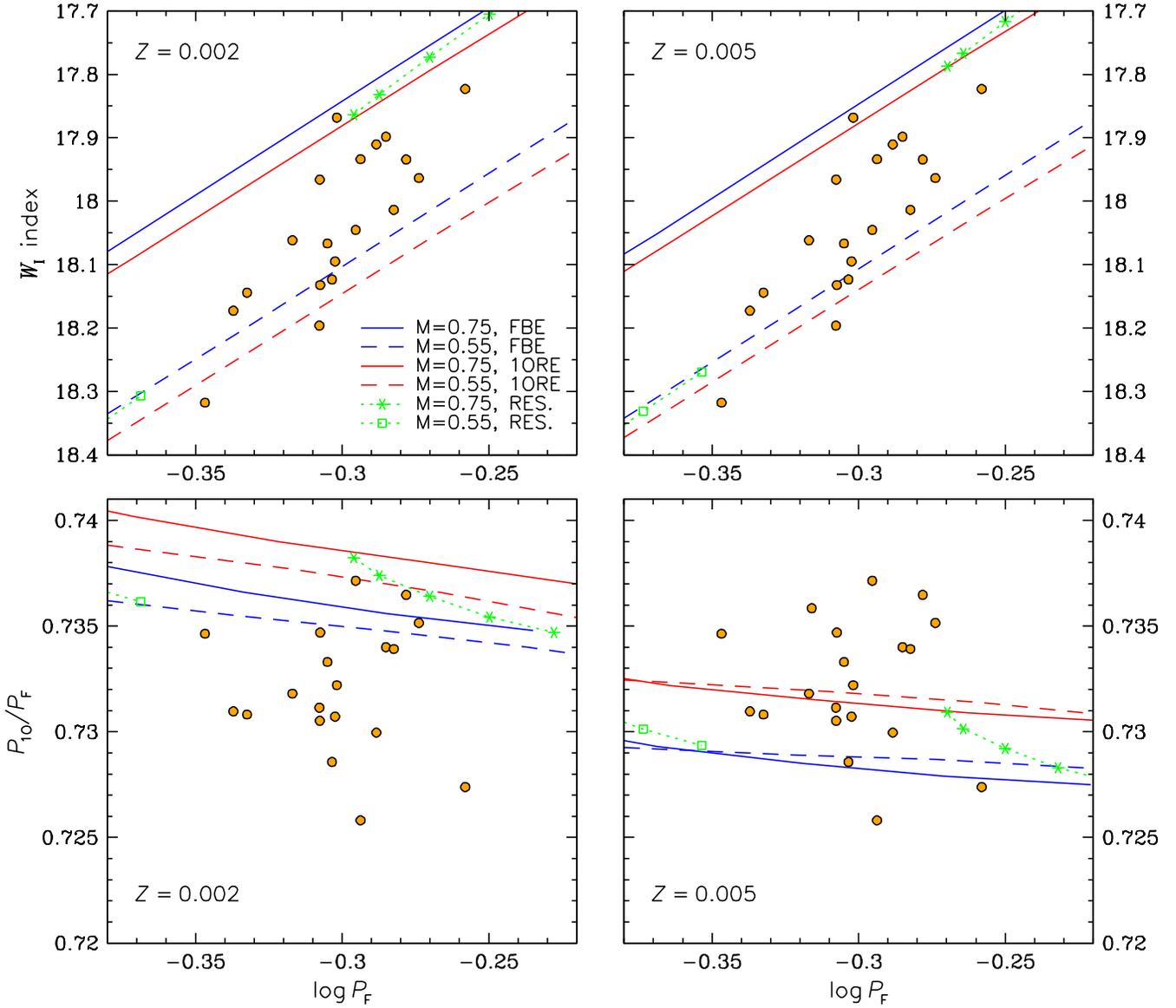}}
\caption{Models in the central part of the instability strip, where the F
and 1O are simultaneously unstable, are confronted with observations. Red
and blue lines connect the corresponding red (1ORE) and blue (FBE) edges
in the range of $L$ chosen to cover the the range of periods of the
anomalous RRd stars. The green lines connect the resonance centers
($\Delta\omega=0$, RES).}
\label{fig:models}
\end{figure*}

The model lines presented in Fig.~\ref{fig:models} correspond to the blue
and red edge of the classical IS. For a given mass, these were delineated
through scan in the model's absolute luminosity, to fully cover the range
of periods observed in anomalous RRd stars.

The comparison of model lines with the observational points in the $(\log
P_{\rm F}, W_I)$ diagrams in the top panels of Fig.~\ref{fig:models}
shows, almost independently of $Z$ adopted in the models, that masses of
the anomalous RRd are spread in the $(0.55-0.75)M_\odot$ range. On the
other hand the comparison in the Petersen diagrams in the bottom panels of
Fig.~\ref{fig:models} shows, almost independently of $M$ adopted in the
models, that metallicities of the anomalous RRd stars spread in the $\sim
(0.002-0.005)$ range (${\rm [Fe/H]}\in(-1.0, -0.5)$). It is essential to
note that period--luminosity diagram and Petersen diagram allow us to
extract orthogonal information about RRd stars, about their masses and
metallicities, respectively.

Pulsation models also provide a hint, what is the cause of anomalous beat
pulsation. In Fig.~\ref{fig:models}, green symbols connected with dotted
lines mark the loci of the parametric resonance, $\Delta\omega=2\omega_{\rm
 1O}-\omega_{\rm F}-\omega_{\rm 2O}=0$. It is clear that one can easily
adjust $M$ and $Z$ so $\Delta\omega$ is close to zero for the majority of
anomalous RRd stars. This suggests that anomalous pulsation form is related
to this resonance. A convenient, qualitative analysis of resonant effects
is possible thanks to amplitude equations formalism, see \cite{dzi82}, on
which we base the below deliberation. Consider an RR~Lyrae star evolving
redward along the HB and pulsating in the 1O mode. When the amplitude of
the relative radius variations, $\epsilon_{\rm 1O,2}$, with the harmonic
frequency, $2\omega_{\rm 1O}$, exceeds certain critical value,
$\epsilon_{\rm crit}$, the F and 2O modes become unstable to the parametric
excitation. The critical amplitude depends on the distance to resonance
center, $\Delta\omega=2\omega_{\rm 1O}-\omega_{\rm F}-\omega_{\rm 2O}$, the
linear damping rates, $-\gamma_{\rm F}$ and $-\gamma_{\rm 2O}$, modified by
1O, and the rate of energy transfer from the ``mother'' mode, 1O, to the
two ``daughter'' modes, F and 2O, which is described by the the factor
$C_{\rm F,2O}$ in the expression for the critical amplitude \citep[see
  e.g.][]{dzi82}:
$$\epsilon_{\rm crit}\equiv\sqrt{\left[1+\left(\frac{\Delta\omega}{\gamma_{\rm F}+\gamma_{\rm 2O}}\right)^2\right]\frac{\gamma_{\rm F}\gamma_{\rm 2O}}{|C_{\rm F,2O}|^2}}.$$

During the evolution toward the red edge of the 1O instability range,
$|\gamma_{\rm F}|$ decreases tending to zero, whereas $|\gamma_{\rm 2O}|$
increases. With the above expression, we may see that the growing disparity
of the damping rates results in widening the frequency range where the
parametric excitation takes place.

\cite{smo07,smo10}, who studied the finite amplitude development of
pulsations in $\beta$~Cep stars and in classical Cepheids with their
nonlinear pulsation code, found that the proximity of the $2\omega_{\rm
 1O}=\omega_{\rm F}+\omega_{\rm 2O}$ resonance is conducive to F+1O
double-mode pulsation. In the frequency spectrum, the 2O mode is hidden due
to its low amplitude and the frequency being exactly the same as that of
the $2\omega_{\rm 1O}-\omega_{\rm F}$ combination peak, which is a result
the phase locking effect.

Blaming the unusual form of pulsation of our anomalous RRd stars to the
same resonance may seem strange because the {\it bona fide} mother mode,
first overtone, has nearly always much lower amplitude in the $I$ band than
the F-mode and the signal of its harmonic, which is responsible for the
mode coupling, is not (or barely) detectable. However, the high amplitude
of the fundamental mode may be explained by its low damping rate. If in the
terminal state of the triple-mode pulsation the amplitudes, $\epsilon$, are
constant or nearly constant, then
$$E_{\rm F}\epsilon_{\rm F}^2\approx E_{\rm 2O}\epsilon_{\rm 2O}^2 \approx-\pol E_{\rm 1O}\epsilon_{\rm 1O}^2\,,$$ 
where $E$ is the rate of energy gain (if $>0$) or loss (if $<0$). These
relations follow from the energy balance and from the symmetry of the
coupling coefficient, $C_{\rm F,2O}$. Thus, we may have
$$ \frac{\epsilon_{\rm F}}{\epsilon_{\rm 1O}}=\sqrt{-\pol\frac{E_{\rm 1O}}{E_{\rm F}}}\gg1.$$

We also note that amplitude equations for the discussed resonance permit
oscillatory solutions around the mean, resonant, triple-mode state. Thus,
the resonance is a plausible explanation for the modulation observed in the
majority of anomalous RRd stars.

\section{Discussion and Conclusions}

Anomalous RRd stars in the Magellanic Clouds are not the first such objects
known in the Universe. A number of double-mode RR~Lyrae stars in the Galactic
bulge have period ratios below 0.742 \citep{sos14}, although not
all of them are considered anomalous. In the bulge, the sequence followed
by classical RRd stars in the Petersen diagram is not truncated at $\approx
0.742$ as in the LMC, but continues towards the lower period ratios (bottom
left panel of Fig.~\ref{fig:pet}). In addition, there is a group outside of
this sequence with distinct features, like the presence of the Blazhko
effect \citep{smo15}, peculiar light curve shape, or anomalous amplitude
ratio. The same systematic differences show four anomalous RRd stars
identified in the globular cluster M3 by \cite{jur14,jur15} (bottom
right panel of Fig.~\ref{fig:pet}). However, anomalous RRd variables in the
Magellanic Clouds constitute the largest sample of such objects known in
any stellar system. This is the first time that there is no doubt that
anomalous RRd stars are not just extreme cases of ``classical'' RRd stars,
but they represent a separate subclass of double-mode RR~Lyrae stars.

Modelling of RR~Lyrae stars {\it ab initio} still represents a considerable
challenge. The problem is particularly difficult for the high metallicity
objects, such as considered in this paper. At the highest metallicities
($Z=0.005$), the main sequence evolutionary time for a star arriving at the
horizontal branch (HB) with $\sim0.55\MS$, exceeds the age of the Universe
\citep[even assuming large mass loss along the red giant branch,
  e.g.][]{pie06,cho16}. At the low metallicities ($Z=0.002$) the stars do
not reach the instability strip after helium ignition, becoming red clump
objects on the cool side of the classical instability strip. The clear
separation of the anomalous RRd stars from the classical RRd sequence in
the Petersen diagram (Fig.~\ref{fig:pet}) suggests a different evolution
channel towards the instability strip, likely through a stripped red giant
phase. We find support for this idea by comparison of the Petersen diagrams
for the LMC and the Galactic bulge. We note a strong disparity in the
incidence rate of anomalous RRd stars in these two environments: as
compared to classical RRd stars, anomalous variables are about 7 times more
numerous in the Galactic bulge than in the LMC. The absence of luminous red
giants has been noted in the inner parts of the Galactic bulge
\citep[e.g.][and references therein]{bar10,kie16}. It has been explained as a
result of interactions of stars in red giant phase with the accretion disc
surrounding the central black hole \citep{kie16}. The stripping may also
result from star-to-star interactions.

We have pointed in the previous section that the models for anomalous RRd
star are close to the parametric resonance, $2\omega_{\rm 1O}=\omega_{\rm
F}+\omega_{\rm 2O}$. One of the observed anomalies in the discussed stars
is the higher amplitude of the fundamental mode. We have shown, based on
the amplitude equations formalism, that it is possible provided the
fundamental mode is weakly damped, but it has to be checked with direct
hydrodynamic modelling. Such modelling should also gain more insight into
the nature of the pronounced bumps visible near the minima of brightness in
some fundamental-mode light curves. Such features may be related to shock
phenomena or arise from the resonant mode interaction. We plan to perform
the nonlinear modeling with the Warsaw code. The spectroscopic confirmation
of high-metallicity of anomalous RRd stars, deduced from pulsation models,
is also needed. We note that we cannot base on the photometric method of
estimating RR~Lyrae stars' metallicity \citep[e.g.][]{jur96}, as the
fundamental mode light curves in anomalous RRd stars have peculiar shape.

\section*{Acknowledgements}

This work has been supported by the Polish Ministry of Science and Higher
Education through the program ``Ideas Plus'' award No. IdP2012 000162. The
OGLE project has received funding from the Polish National Science Centre
grant MAESTRO no. 2014/14/A/ST9/00121. RS acknowledges support from the
Polish National Science Centre grants no. DEC-2012/05/B/ST9/03932 and
DEC-2015/17/B/ST9/03421.


\appendix
\section{Complete light-curve solutions}

\begin{table*}
\centering
\caption{Light curve solution for OGLE-LMC-RRLYR-00013. The consecutive columns contain frequency id., frequency value, amplitude with standard error, phase with standard error and remarks. Entries are sorted by increasing frequency. In case of frequency values the error of the last two digits of independent frequencies is given in parenthesis. `bl' in the `remarks' column indicates a frequency of the Blazhko modulation in case no significant peak was detected in the frequency spectrum directly at $\nu=\fb$. Amplitude and phase are not given then.}
\begin{tabular}{lr@{.}lccccc}
freq. id & \multicolumn{2}{c}{$f$\thinspace [d$^{-1}$]}& $A$\thinspace [mmag] & $\sigma$ & $\phi$ [rad] & $\sigma$ & remarks\\
\hline
 $\fb$          &  0&008077(18)  &      &     &        &      & bl\\
 $\ff-\fb$      &  2&010188      &  91  &  4  &  5.93  & 0.57 &\\
 $\ff$          &  2&018264(12)  &  87  &  4  &  3.38  & 0.46 &\\
 $\fo-\fb$      &  2&744229      &  29  &  4  &  3.07  & 0.88 &\\
 $\fo$          &  2&752306(21)  &  59  &  4  &  3.96  & 0.78 &\\
 $2\ff$         &  4&036529      &  28  &  4  &  2.33  & 0.93 &\\
\hline
\end{tabular}
\end{table*}

\begin{table*}
\centering
\caption{Same as Tab.~1 for OGLE-LMC-RRLYR-03878.}
\begin{tabular}{lr@{.}lccccc}
freq. id & \multicolumn{2}{c}{$f$\thinspace [d$^{-1}$]}& $A$\thinspace [mmag] & $\sigma$ & $\phi$ [rad] & $\sigma$ & remarks\\
\hline
 $\fb$          &  0&004964(22)  &      &     &        &      & bl\\
 $\ff-\fb$      &  1&873496      &  59  &  3  &  0.81  & 0.65 &\\
 $\ff$          &  1&878460(12)  &  62  &  3  &  5.53  & 0.46 &\\
 $\fo$          &  2&555219(17)  &  63  &  3  &  0.80  & 0.64 &\\
 $2\ff$         &  3&756919      &  31  &  3  &  0.23  & 0.91 &\\
\hline
\end{tabular}
\end{table*}

\begin{table*}
\centering
\caption{Same as Tab.~1 for OGLE-LMC-RRLYR-04176.}
\begin{tabular}{lr@{.}lccccc}
freq. id & \multicolumn{2}{c}{$f$\thinspace [d$^{-1}$]}& $A$\thinspace [mmag] & $\sigma$ & $\phi$ [rad] & $\sigma$ & remarks\\
\hline
 $\fb$          &  0&05070(3)    &      &     &        &      & bl\\
 $\ff-\fb$      &  1&923317      &  40  &  4  &  5.05  & 1.19 &\\
 $\ff$          &  1&974019(8)   & 152  &  4  &  4.20  & 0.29 &\\
 $\fo$          &  2&67790(4)    &  33  &  4  &  3.54  & 1.48 &\\
 $2\ff$         &  3&948039      &  42  &  4  &  4.05  & 0.58 &\\
\hline
\end{tabular}
\end{table*}

\begin{table*}
\centering
\caption{Same as Tab.~1 for OGLE-LMC-RRLYR-07276.}
\begin{tabular}{lr@{.}lccccc}
freq. id & \multicolumn{2}{c}{$f$\thinspace [d$^{-1}$]}& $A$\thinspace [mmag] & $\sigma$ & $\phi$ [rad] & $\sigma$ & remarks\\
\hline
 $\fb$          &  0&022267(21)  &      &     &        &      & bl\\
 $\ff-\fb$      &  2&007461      &  62  &  4  &  1.92  & 0.72 &\\
 $\ff$          &  2&029729(8)   & 118  &  4  &  3.52  & 0.28 &\\
 $\ff+\fb$      &  2&051996      &  23  &  4  &  2.72  & 0.96 &\\
 $\fo$          &  2&762746(21)  &  58  &  4  &  3.40  & 0.79 &\\
 $2\ff$         &  4&059457      &  46  &  4  &  2.19  & 0.58 &\\
 $\ff+\fo-\fb$  &  4&770207      &  28  &  4  &  1.17  & 1.07 &\\
 $\ff+\fo+\fb$  &  4&814742      &   7  &  4  &  4.98  & 1.32 &\\
 $3\ff$         &  6&089186      &  20  &  4  &  1.80  & 0.88 &\\
\hline
\end{tabular}
\end{table*}

\begin{table*}
\centering
\caption{Same as Tab.~1 for OGLE-LMC-RRLYR-08767.}
\begin{tabular}{lr@{.}lccccc}
freq. id & \multicolumn{2}{c}{$f$\thinspace [d$^{-1}$]}& $A$\thinspace [mmag] & $\sigma$ & $\phi$ [rad] & $\sigma$ & remarks\\
\hline
 $\fb$          &  0&004477(24)  &      &     &        &      & bl\\
 $\ff-\fb$      &  1&911232      &  49  &  4  &  4.77  & 0.92 &\\
 $\ff$          &  1&915709(9)   & 106  &  4  &  3.28  & 0.33 &\\
 $\fo$          &  2&610255(18)  &  59  &  4  &  5.85  & 0.69 &\\
 $2\ff$         &  3&831418      &  39  &  4  &  1.71  & 0.67 &\\
\hline
\end{tabular}
\end{table*}

\begin{table*}
\centering
\caption{Same as Tab.~1 for OGLE-LMC-RRLYR-08917.}
\begin{tabular}{lr@{.}lccccc}
freq. id & \multicolumn{2}{c}{$f$\thinspace [d$^{-1}$]}& $A$\thinspace [mmag] & $\sigma$ & $\phi$ [rad] & $\sigma$ & remarks\\
\hline
 $\fb$          &  0&006103(8)   &      &     &        &      & bl\\
 $\ff$          &  2&011122(11)  &  85  &  3  &  6.12  & 0.41 &\\
 $\ff+2\fb$     &  2&023329      &  55  &  3  &  4.16  & 0.57 &\\
 $\fo-\fb$      &  2&754261      &  69  &  3  &  2.90  & 0.47 &\\
 $\fo$          &  2&760365(12)  &  55  &  3  &  1.66  & 0.44 &\\
 $\fo+\fb$      &  2&766468      &  52  &  3  &  3.92  & 0.60 &\\
 $\fu$          &  4&02089(7)    &  18  &  3  &  5.98  & 2.52 &\\
 $2\ff+2\fb$    &  4&034452      &  31  &  3  &  5.71  & 0.77 &\\
 $\ff+\fo-\fb$  &  4&765384      &  20  &  3  &  5.31  & 0.69 &\\
 $\ff+\fo+\fb$  &  4&777590      &  17  &  3  &  5.10  & 0.61 &\\
\hline
\end{tabular}
\end{table*}

\begin{table*}
\centering
\caption{Same as Tab.~1 for OGLE-LMC-RRLYR-09866.}
\begin{tabular}{lr@{.}lccccc}
freq. id & \multicolumn{2}{c}{$f$\thinspace [d$^{-1}$]}& $A$\thinspace [mmag] & $\sigma$ & $\phi$ [rad] & $\sigma$ & remarks\\
\hline
 $\fb$          &  0&00494(3)    &      &     &        &      & bl\\
 $\fo-\ff-\fb$  &  0&734274      &  10  &  3  &  0.66  & 1.83 &\\
 $\fo-\ff$      &  0&739217      &  22  &  3  &  3.93  & 1.15 &\\
 $\ff-\fb$      &  2&001210      &  53  &  3  &  4.54  & 1.22 &\\
 $\ff$          &  2&006152(12)  &  66  &  3  &  3.10  & 0.45 &\\
 $\ff+\fb$      &  2&011095      &  17  &  3  &  4.62  & 1.41 &\\
 $\fo$          &  2&745369(27)  &  50  &  3  &  0.74  & 1.01 &\\
 $2\ff$         &  4&012305      &  58  &  3  &  1.33  & 0.90 &\\
\hline
\end{tabular}
\end{table*}

\begin{table*}
\centering
\caption{Same as Tab.~1 for OGLE-LMC-RRLYR-10802.}
\begin{tabular}{lr@{.}lccccc}
freq. id & \multicolumn{2}{c}{$f$\thinspace [d$^{-1}$]}& $A$\thinspace [mmag] & $\sigma$ & $\phi$ [rad] & $\sigma$ & remarks\\
\hline
 $\fb$          &  0&019561(15)  &  26  &  3  &  0.13  & 0.59 &\\
 $\ff-\fb$      &  1&983981      &  60  &  3  &  1.42  & 0.53 &\\
 $\ff$          &  2&003543(7)   &  90  &  3  &  1.44  & 0.25 &\\
 $\fo$          &  2&73627(6)    &  18  &  3  &  5.02  & 2.09 &\\
 $2\ff$         &  4&007086      &  64  &  3  &  4.11  & 0.50 &\\
 $3\ff$         &  6&010628      &  20  &  3  &  1.58  & 0.77 &\\
 $4\ff$         &  8&014171      &  13  &  3  &  4.86  & 0.99 &\\
\hline
\end{tabular}
\end{table*}

\begin{table*}
\centering
\caption{Same as Tab.~1 for OGLE-LMC-RRLYR-12487.}
\begin{tabular}{lr@{.}lccccc}
freq. id & \multicolumn{2}{c}{$f$\thinspace [d$^{-1}$]}& $A$\thinspace [mmag] & $\sigma$ & $\phi$ [rad] & $\sigma$ & remarks\\
\hline
 $\fb$          &  0&00765(4)    &      &     &        &      & bl\\
 $\fo-\ff$      &  0&802569      &  20  &  4  &  0.89  & 0.98 &\\
 $\ff-\fb$      &  2&214378      &  32  &  4  &  0.71  & 1.36 &\\
 $\ff$          &  2&222033(7)   & 173  &  4  &  5.00  & 0.26 &\\
 $\fo$          &  3&024602(24)  &  50  &  4  &  6.16  & 0.92 &\\
 $\fo+\fb$      &  3&032257      &  22  &  4  &  1.51  & 1.48 &\\
 $2\ff$         &  4&444067      &  56  &  4  &  5.67  & 0.53 &\\
\hline
\end{tabular}
\end{table*}

\begin{table*}
\centering
\caption{Same as Tab.~1 for OGLE-LMC-RRLYR-14584.}
\begin{tabular}{lr@{.}lccccc}
freq. id & \multicolumn{2}{c}{$f$\thinspace [d$^{-1}$]}& $A$\thinspace [mmag] & $\sigma$ & $\phi$ [rad] & $\sigma$ & remarks\\
\hline
 $\fb$          &  0&004880(20)  &      &     &        &      & bl\\
 $\fu$          &  0&81330(6)    &  28  &  5  &  3.48  & 2.20 &\\
 $\ff-\fb$      &  2&065133      &  54  &  5  &  3.42  & 0.74 &\\
 $\ff$          &  2&070013(9)   & 155  &  5  &  5.37  & 0.34 &\\
 $\ff+\fb$      &  2&074893      &  50  &  5  &  3.63  & 0.93 &\\
 $\fo$          &  2&81301(3)    &  51  &  5  &  0.74  & 1.17 &\\
 $\fo+\fb$      &  2&817892      &  47  &  5  &  6.07  & 1.16 &\\
 $2\ff$         &  4&140025      &  33  &  5  &  6.20  & 0.68 &\\
\hline
\end{tabular}
\end{table*}

\begin{table*}
\centering
\caption{Same as Tab.~1 for OGLE-LMC-RRLYR-19077.}
\begin{tabular}{lr@{.}lccccc}
freq. id & \multicolumn{2}{c}{$f$\thinspace [d$^{-1}$]}& $A$\thinspace [mmag] & $\sigma$ & $\phi$ [rad] & $\sigma$ & remarks\\
\hline
 $\fb$          &  0&01773(3)    &      &     &        &      & bl\\
 $\ff$          &  2&031038(17)  &  73  &  4  &  4.70  & 0.64 &\\
 $\fo-\fb$      &  2&760208      &  44  &  4  &  3.01  & 0.97 &\\
 $\fo$          &  2&777934(22)  &  59  &  4  &  2.82  & 0.83 &\\
 $\ff+\fo-\fb$  &  4&791246      &  25  &  4  &  3.41  & 1.12 &\\
\hline
\end{tabular}
\end{table*}

\begin{table*}
\centering
\caption{Same as Tab.~1 for OGLE-LMC-RRLYR-21363.}
\begin{tabular}{lr@{.}lccccc}
freq. id & \multicolumn{2}{c}{$f$\thinspace [d$^{-1}$]}& $A$\thinspace [mmag] & $\sigma$ & $\phi$ [rad] & $\sigma$ & remarks\\
\hline
 $\ff$          &  2&030567(35)  &  45  &  5  &  1.03  & 1.34 &\\
 $\fo$          &  2&779582(18)  &  93  &  5  &  5.85  & 0.67 &\\
\hline
\end{tabular}
\end{table*}

\begin{table*}
\centering
\caption{Same as Tab.~1 for OGLE-LMC-RRLYR-22167.}
\begin{tabular}{lr@{.}lccccc}
freq. id & \multicolumn{2}{c}{$f$\thinspace [d$^{-1}$]}& $A$\thinspace [mmag] & $\sigma$ & $\phi$ [rad] & $\sigma$ & remarks\\
\hline
 $\fb$          &  0&006980(15)  &      &     &        &      & bl\\
 $\ff-\fb$      &  1&935345      &  58  &  4  &  1.48  & 0.63 &\\
 $\ff$          &  1&942325(13)  &  53  &  4  &  1.19  & 0.50 &\\
 $\ff+\fb$      &  1&949305      &  46  &  4  &  4.93  & 0.83 &\\
 $\fo-2\fb$     &  2&646871      &  14  &  4  &  4.74  & 1.16 &\\
 $\fo-\fb$      &  2&653851      &  16  &  4  &  1.71  & 0.77 &\\
 $\fo$          &  2&660831(16)  &  74  &  4  &  0.01  & 0.58 &\\
 $\ff+\fo-\fb$  &  4&596176      &  21  &  4  &  3.69  & 0.87 &\\
\hline
\end{tabular}
\end{table*}

\begin{table*}
\centering
\caption{Same as Tab.~1 for OGLE-LMC-RRLYR-24495.}
\begin{tabular}{lr@{.}lccccc}
freq. id & \multicolumn{2}{c}{$f$\thinspace [d$^{-1}$]}& $A$\thinspace [mmag] & $\sigma$ & $\phi$ [rad] & $\sigma$ & remarks\\
\hline
 $\fb$          &  0&005737(27)  &      &     &        &      & bl\\
 $\ff-\fb$      &  1&921847      &  37  &  4  &  1.02  & 0.98 &\\
 $\ff$          &  1&927583(5)   & 167  &  4  &  1.44  & 0.20 &\\
 $\fo$          &  2&626123(29)  &  34  &  4  &  2.66  & 1.09 &\\
 $2\ff$         &  3&855167      &  53  &  4  &  4.75  & 0.40 &\\
 $\ff+\fo-\fb$  &  4&547969      &  16  &  4  &  5.31  & 1.23 &\\
 $3\ff$         &  5&782750      &  32  &  4  &  2.37  & 0.60 &\\
 $2\ff+\fo-\fb$ &  6&475553      &  20  &  4  &  3.01  & 1.21 &\\
 $4\ff$         &  7&710334      &  17  &  4  &  5.99  & 0.83 &\\
 $3\ff+\fo-\fb$ &  8&403136      &  17  &  4  &  0.80  & 1.24 &\\
\hline
\end{tabular}
\end{table*}

\begin{table*}
\centering
\caption{Same as Tab.~1 for OGLE-LMC-RRLYR-27200.}
\begin{tabular}{lr@{.}lccccc}
freq. id & \multicolumn{2}{c}{$f$\thinspace [d$^{-1}$]}& $A$\thinspace [mmag] & $\sigma$ & $\phi$ [rad] & $\sigma$ & remarks\\
\hline
 $\fb$          &  0&004957(17)  &      &     &        &      & bl\\
 $\ff-\fb$      &  1&806415      &  54  &  3  &  4.63  & 0.63 &\\
 $\ff$          &  1&811372(6)   & 103  &  3  &  1.33  & 0.22 &\\
 $\ff+\fb$      &  1&816328      &  12  &  3  &  5.87  & 0.76 &\\
 $\fo$          &  2&49025(4)    &  21  &  3  &  0.50  & 1.63 &\\
 $2\ff$         &  3&622743      &  51  &  3  &  4.22  & 0.43 &\\
 $3\ff$         &  5&434115      &  15  &  3  &  2.39  & 0.67 &\\
\hline
\end{tabular}
\end{table*}

\begin{table*}
\centering
\caption{Same as Tab.~1 for OGLE-LMC-RRLYR-27329.}
\begin{tabular}{lr@{.}lccccc}
freq. id & \multicolumn{2}{c}{$f$\thinspace [d$^{-1}$]}& $A$\thinspace [mmag] & $\sigma$ & $\phi$ [rad] & $\sigma$ & remarks\\
\hline
 $\fb$          &  0&02229(4)    &      &     &        &      & bl\\
 $\ff-\fb$      &  1&874958      &  40  &  5  &  4.69  & 1.55 &\\
 $\ff$          &  1&897251(17)  &  96  &  5  &  2.27  & 0.66 &\\
 $\fo$          &  2&576085(28)  &  59  &  5  &  6.23  & 1.05 &\\
\hline
\end{tabular}
\end{table*}

\begin{table*}
\centering
\caption{Same as Tab.~1 for OGLE-LMC-RRLYR-30248.}
\begin{tabular}{lr@{.}lccccc}
freq. id & \multicolumn{2}{c}{$f$\thinspace [d$^{-1}$]}& $A$\thinspace [mmag] & $\sigma$ & $\phi$ [rad] & $\sigma$ & remarks\\
\hline
 $\ff$          &  2&149873(6)   & 139  &  4  &  4.58  & 0.21 &\\
 $\fo$          &  2&94174(4)    &  34  &  4  &  1.10  & 1.28 &\\
 $2\ff$         &  4&299747      &  51  &  4  &  4.56  & 0.43 &\\
 $3\ff$         &  6&449621      &  22  &  4  &  5.28  & 0.66 &\\
 $4\ff$         &  8&599494      &  20  &  4  &  5.39  & 0.87 &\\
\hline
\end{tabular}
\end{table*}

\begin{table*}
\centering
\caption{Same as Tab.~1 for OGLE-LMC-RRLYR-31101.}
\begin{tabular}{lr@{.}lccccc}
freq. id & \multicolumn{2}{c}{$f$\thinspace [d$^{-1}$]}& $A$\thinspace [mmag] & $\sigma$ & $\phi$ [rad] & $\sigma$ & remarks\\
\hline
 $\fb$          &  0&004851(16)  &      &     &        &      & bl\\
 $\ff-\fb$      &  2&168182      &  54  &  4  &  3.62  & 0.63 &\\
 $\ff$          &  2&173033(8)   &  90  &  4  &  4.24  & 0.29 &\\
 $\fo-2\fb$     &  2&963147      &  23  &  4  &  4.57  & 1.37 &\\
 $\fo$          &  2&972848(22)  &  52  &  4  &  4.61  & 0.82 &\\
 $2\ff$         &  4&346066      &  58  &  4  &  3.88  & 0.57 &\\
\hline
\end{tabular}
\end{table*}

\begin{table*}
\centering
\caption{Same as Tab.~1 for OGLE-LMC-RRLYR-37673.}
\begin{tabular}{lr@{.}lccccc}
freq. id & \multicolumn{2}{c}{$f$\thinspace [d$^{-1}$]}& $A$\thinspace [mmag] & $\sigma$ & $\phi$ [rad] & $\sigma$ & remarks\\
\hline
 $\fo-\ff$      &  0&742925      &  16  &  4  &  5.96  & 0.87 &\\
 $\ff$          &  1&966541(10)  & 107  &  4  &  5.81  & 0.39 &\\
 $\fo$          &  2&709466(19)  &  61  &  4  &  0.09  & 0.70 &\\
 $\ff+\fo$      &  4&676008      &  16  &  4  &  1.94  & 0.79 &\\
\hline
\end{tabular}
\end{table*}

\begin{table*}
\centering
\caption{Same as Tab.~1 for OGLE-LMC-RRLYR-38132.}
\begin{tabular}{lr@{.}lccccc}
freq. id & \multicolumn{2}{c}{$f$\thinspace [d$^{-1}$]}& $A$\thinspace [mmag] & $\sigma$ & $\phi$ [rad] & $\sigma$ & remarks\\
\hline
 $\ff$          &  2&074566(19)  &  95  &  6  &  1.60  & 0.73 &\\
 $\fo$          &  2&834866(19)  &  87  &  6  &  3.24  & 0.77 &\\
\hline
\end{tabular}
\end{table*}

\begin{table*}
\centering
\caption{Same as Tab.~1 for OGLE-SMC-RRLYR-1125.}
\begin{tabular}{lr@{.}lccccc}
freq. id & \multicolumn{2}{c}{$f$\thinspace [d$^{-1}$]}& $A$\thinspace [mmag] & $\sigma$ & $\phi$ [rad] & $\sigma$ & remarks\\
\hline
 $\fb$          &  0&01343(7)    &      &     &        &      & bl\\
 $\ff-\fb$      &  2&089900      &  32  &  6  &  4.73  & 2.48 &\\
 $\ff$          &  2&103327(10)  & 166  &  7  &  3.02  & 0.41 &\\
 $\fo$          &  2&85274(4)    &  52  &  6  &  4.42  & 1.46 &\\
 $2\ff$         &  4&206654      &  45  &  6  &  1.83  & 0.82 &\\
\hline
\end{tabular}
\end{table*}

\begin{table*}
\centering
\caption{Same as Tab.~1 for OGLE-SMC-RRLYR-4726.}
\begin{tabular}{lr@{.}lccccc}
freq. id & \multicolumn{2}{c}{$f$\thinspace [d$^{-1}$]}& $A$\thinspace [mmag] & $\sigma$ & $\phi$ [rad] & $\sigma$ & remarks\\
\hline
 $\fb$          &  0&00692(6)    &      &     &        &      & bl\\
 $\ff-\fb$      &  1&906728      &  65  &  9  &  5.63  & 2.36 &\\
 $\ff$          &  1&913648(18)  & 185  &  8  &  1.51  & 0.74 &\\
 $\fo$          &  2&63220(9)    &  41  &  9  &  1.18  & 4.01 &\\
 $2\ff$         &  3&827297      &  46  &  8  &  4.74  & 1.49 &\\
\hline
\end{tabular}
\end{table*}

\bsp	
\label{lastpage}
\end{document}